\documentclass[12pt,letterpaper]{article}

\usepackage{jheppub}
\usepackage{color}
\usepackage{graphicx}
\usepackage{wrapfig,enumerate,slashed}
\usepackage{tabularx} 

\usepackage{amsmath}
\usepackage{wasysym} 
\usepackage{graphicx}
\usepackage{color}
\usepackage{hyperref}

\newcommand{\be}{\begin{eqnarray}}
\newcommand{\ee}{\end{eqnarray}}

\newcommand{\bee}{\begin{eqnarray}}
\newcommand{\eee}{\end{eqnarray}}
\newcommand{\beeq}{\begin{equation}}
\newcommand{\eeeq}{\end{equation}}

\begin{document}

\title{Searching new physics in rare $B$-meson decays into multiple muons}

\author[a]{Mikael Chala,}
\author[b,c]{Ulrik Egede}
\author[a]{and Michael Spannowsky}

\affiliation[a]{Institute for Particle Physics Phenomenology, Department of Physics, Durham University, Durham, DH1 3LE, UK}

\affiliation[b]{Imperial College London, London, United Kingdom}
\affiliation[b]{School of Physics and Astronomy, Monash University, Melbourne, Australia}

\emailAdd{mikael.chala@durham.ac.uk}
\emailAdd{ulrik.egede@monash.edu}
\emailAdd{michael.spannowsky@durham.ac.uk}

\abstract{New heavy vector bosons and light scalars are 
predicted in a plethora of models of new physics. In particular, in new strongly 
interacting sectors they play the role of the $\rho$ and $\pi$ mesons in QCD. We 
show that some of their interactions, for example those required for the 
explanation of the $B$ anomalies and the $g-2$ of the muon, can be only probed 
in $B$ meson decays. We highlight new golden channels not yet studied 
experimentally, including $B^+ \rightarrow K^+(D^+)\mu^+\mu^-\mu^+\mu^-$ and 
$B^0\rightarrow K^{*0}\mu^+\mu^-\mu^+\mu^-$. Relying on generator level simulations for 
data taking with the LHCb detector, we determine the reach 
of 
this facility to the aforementioned processes. We show that branching ratios as 
small as $9\times 10^{-12}$ ($3.2\times 10^{-10}$) and $2.7\times 10^{-11}$ can 
be tested at 
the $95\, \%$ CL respectively.
}

\preprint{IPPP/19/12}

\maketitle


\section{Introduction}
\label{sec:intro}

New heavy vector bosons $V$ and light scalars $a$ are common predictions of 
different scenarios of 
physics beyond the Standard Model (SM). The former appear in extensions of the 
SM 
gauge group, including theories of grand unification~\cite{Georgi:1974sy,Georgi:1974yf,Pati:1974yy,GellMann:1976pg,Langacker:1980js} and string 
constructions~\cite{Hewett:1988xc}. They are also natural in composite sectors~\cite{Weinberg:1975gm,Susskind:1978ms,Farhi:1980xs,Kaplan:1983sm} and 
their holographic relatives~\cite{ArkaniHamed:2000ds,Rattazzi:2000hs}. Recently, 
new vectors 
at the TeV scale have been also proposed as a plausible 
explanation~\cite{DAmico:2017mtc} of the 
anomalies observed in the branching fractions, angular distributions and lepton universality tests of the decays $B^{+(0)}\rightarrow K^{+(*)}\ell^+\ell^-$~\cite{Aaij:2013qta,Aaij:2014ora,Aaij:2014pli,Aaij:2015esa,Aaij:2015oid,
Aaij:2017vbb,Wehle:2016yoi, Khachatryan:2015isa,Sirunyan:2017dhj}. Models 
prediciting ultralight scalars $a$ have been studied~\cite{Bauer:2017ris} for 
collider phenomenology. 
Likewise, they have been also studied in light of the observed disagreement 
between the predicted and the observed values	
of the anomalous magnetic moment of the muon~\cite{Bauer:2017nlg,Liu:2018xkx}.

Perhaps, the most traditional scenario involving both new scalars and
vectors consists of a new strongly interacting sector extending the
SM. In this case, $V$ and $a$ play the role of the $\rho$ and $\pi$
mesons in QCD. The separation of these scales is explained by the
pseudo-Nambu Goldstone boson nature of the latter. We show that rare
$B$ decays can be naturally expected in this context. In the situation
where the scalar decays to two muons, these include
$B^0_{(s)}\rightarrow aa$, which have already been searched for at
LHCb~\cite{Aaij:2016kfs}, as well as $B^{+/0} \rightarrow M aa$ with
$M=K^+, D^+, K^*$.

In this article, we perform simulations to estimate the reach of the 
LHCb experiment to the aforementioned processes with the currently available data as well as with the anticipated upgrades. Our choice of parameters is motivated by 
the $B$ and $(g-2)_\mu$ anomalies. However, our results are of much broader 
applicability. The paper is organized as follows. In 
Section~\ref{sec:lagrangian} we introduce the generic Lagrangian we are 
interested in; we comment on constraints on the different parameters and 
compute the amplitudes for the different $B$ decays. In 
Section~\ref{sec:models}, we match a particular composite Higgs model to the 
Lagrangian above. We 
study the new rare $B$ meson decays in Section~\ref{sec:BF}. We subsequently interpret these results 
in the model introduced before. Finally, we conclude in Section~\ref{sec:conclusions}.

\section{Generic Lagrangian}
\label{sec:lagrangian}
Let us extend the SM with a new vector boson $V$ and a new scalar  
$a$ with masses of the order of TeV and GeV, respectively. They are both 
singlets of the SM gauge group. At energies below the electroweak scale 
$v\sim 246$ GeV, the Lagrangian we are interested in is
\begin{align}\label{eq:Lagrangian}\nonumber
 L &= \frac{1}{2}m_V^2 V_\mu V^\mu + \frac{1}{2}m_a^2 a^2 \\
 &+V_\mu \left[g^u_{ij} (\overline{u_L^i}\gamma^\mu u_L^j) + g^d_{ij} 
(\overline{d_L^i}\gamma^\mu d_L^j) + \text{h.c.}\right]+ g_{ij}^ea 
\left[\overline{l_L^i} e_R^j + \text{h.c.}\right]+g' V_\mu 
(a\partial^\mu a)~.\,
\end{align}
where $u_L^i, d_L^i$ stand for the $i$-th family of left-handed up and down 
quarks, and $l_L^i, e_R^i$ stand for the $i$-th family of left- and 
right-handed leptons, respectively. Inspired by recently observed flavour anomalies, we will 
focus mostly on the case in which the only non-vanishing $g$ coupling is 
$g_{23}^d$. To a lesser extent, we will also consider $g_{13}^d\neq 0$. These 
couplings are constrained by measurements of $\Delta M_s$
and $\Delta M_d$~\cite{Foldenauer:2016rpi}, respectively.
Thus, for $m_V = 1$ TeV, $g_{23}^d \lesssim 0.002$ and $g_{13}^d$ is bounded 
to be about one order 
of magnitude smaller. 

\begin{figure}[t]
 \includegraphics[width=\columnwidth]{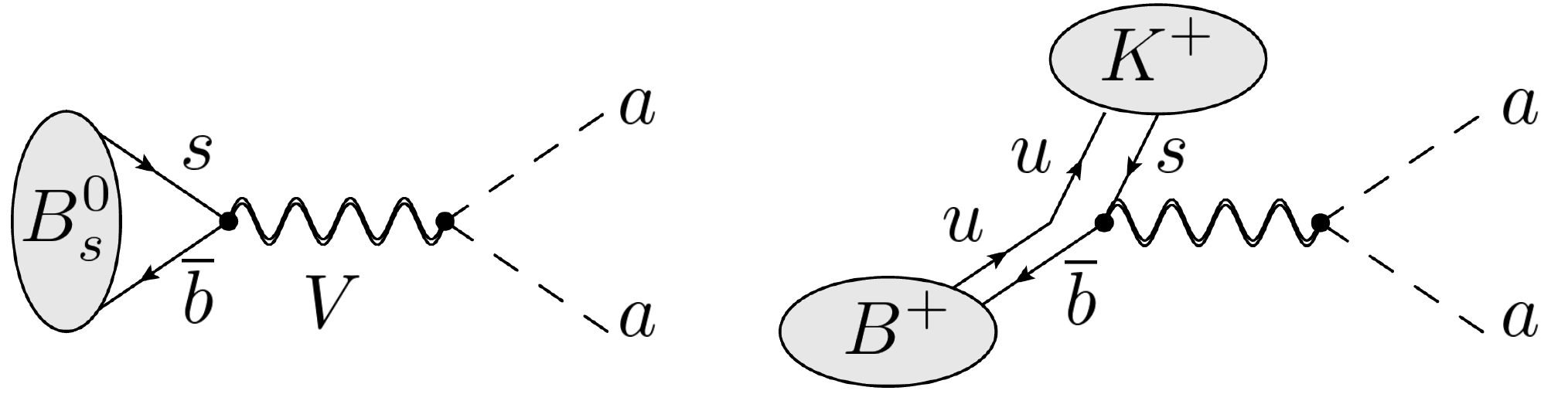}
 \caption{\it Diagrams of the decays $B_s^0 \to aa$ (left) and $B^+ \to K^+ a 
a$ 
(right) mediated by a heavy vector boson $V$.}\label{fig:diagrams}
\end{figure}

Together with a non-vanishing $g'$, these couplings trigger rare $B$ decays as shown in Fig.~\ref{fig:diagrams}.
The decay width for $B_s^0\rightarrow aa$ is
\begin{equation}
 \Gamma = \frac{f_B^2}{32 \pi m_V^4} 
(g_{23}^d g')^2 m_B^3 \sqrt{1-\frac{4 m_a^2}{m_B^2}}~,
\end{equation}
with $f_B \sim 0.23$ GeV~\cite{Cheung:2006tm}. Similar expressions 
hold for other decay modes, \textit{e.g.} $B^0\rightarrow a a$.

The amplitude for $B^+\rightarrow K^+ a a$ reads:
\begin{equation}
 \mathcal{M} = \frac{g_{23}^d g'}{m_V^2} \langle K(p_3)|\overline{s}\gamma_\mu 
b|B(p)\rangle (p_1+p_2)^\mu
\end{equation}
with~\cite{Ball:2004ye}
\begin{equation}
 \langle K(p_3)|\overline{s}\gamma_\mu 
b|B(p)\rangle = f_+^P(q^2) \left[(p+p_3)_\mu - 
\frac{m_{B}^2-m_K^2}{q^2}q_\mu\right] + f_0(q^2)\frac{m_{B}^2-m_K^2}{q^2}q_\mu 
~.
\end{equation}
The transferred momentum is $q^2 = (p-p_3)^2$, and varies 
between $q^2_\text{min} = 4 m_a^2$ and $q^2_\text{max} = 
(m_{B}-m_K)^2$. The contraction of this matrix element with $(p_1+p_2) = 
q$ in the amplitude annihilates the $f_+^P(q^2)$ part. Altogether, we obtain
\begin{equation}
 \frac{d\Gamma}{dq^2} = \frac{(g_{23}^d g')^2}{512\pi^3 m_V^4 m_{B}^3} 
(m_{B}^2-m_K^2)^2 F(q^2)~,
\end{equation}
with
\begin{equation}
 F(q^2)\equiv   \sqrt{m_{B}^4 + 
(m_K^2-q^2)^2 - 2m_{B^+}^2 (m_K^2+q^2)} 
\sqrt{1-\frac{4m_a^2}{q^2}} |f_0(q^2)|^2~.
\end{equation}
In the approximation $m_K,m_a\rightarrow 0$, $f_0(q^2)\rightarrow 1$, one 
easily obtains
\begin{equation}
\Gamma\sim \frac{(g_{23}^d  g')^2}{1024 \pi^3 m_V^4} m_B^5~.
\end{equation}

\begin{figure}[t]
 \hspace{-0.3cm}\includegraphics[width=0.48\columnwidth]{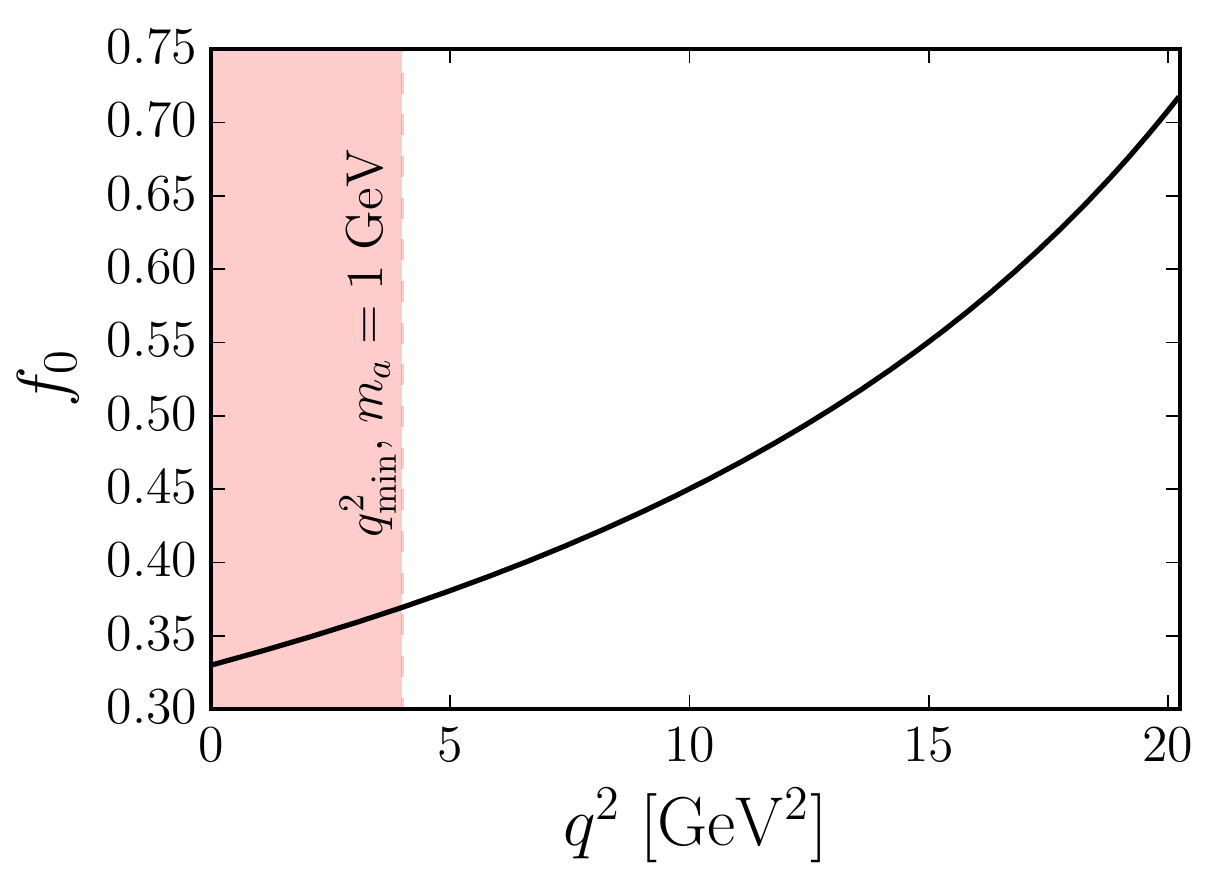}
 \includegraphics[width=0.53\columnwidth]{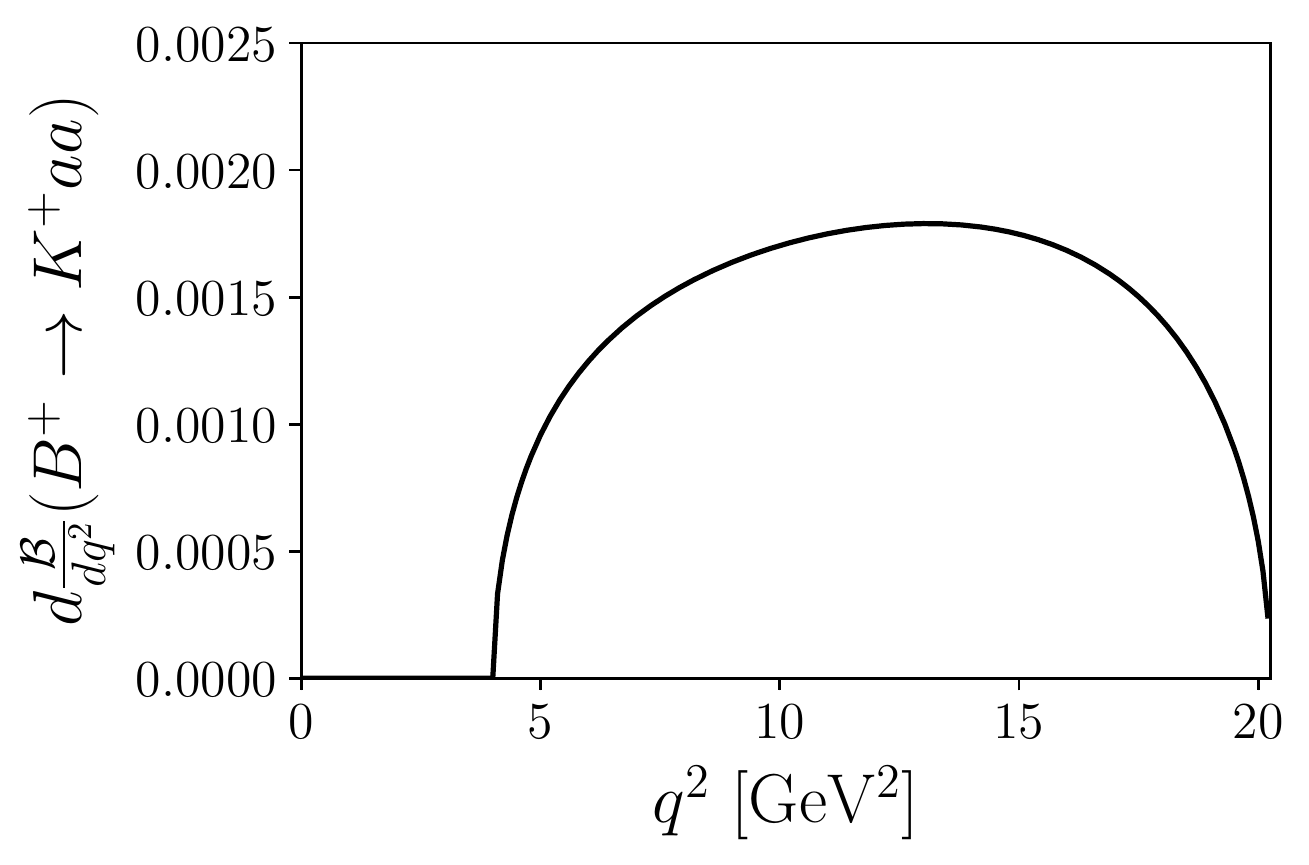} 
 \caption{\it Left) Scalar momentum form factor for $B^+\to K^+$ as a function of the transferred momentum $q^2$. Right) Differential branching ratio for $B^+\rightarrow K^+ a a$ as a function of the momentum transferred $q^2$. In both cases, we have fixed $m_V = 1$ TeV, $m_a = 1$ GeV, $g_{23}=g'=1$. }\label{fig:formfactor}
\end{figure}

Following Ref.~\cite{Ball:2004ye}, we parametrize the form factor as 
$f_0(q^2) = r_2/(1-q^2/m_\text{fit}^2)$, with $r_2 = 0.330$ and 
$m^2_\text{fit} = 37.46$ GeV$^2$; see Fig.~\ref{fig:formfactor}.
Similar expressions hold for other processes, \textit{e.g.}  
$B^0\rightarrow 
K^{*0}a a$ or $B_c^+\to D^+ a a$. The latter is however hard 
to test at the LHCb and we will not consider it. The reason is that the 
$B_c^+$ production cross section is much smaller and the $B_c^+$ width is larger 
(which reduces both the impact of the new interactions and the experimental 
efficiency).

We note also that final states containing one meson and $a a$
 can probe effective operators containing four quarks 
and two
light scalars
(12 of these operators are present in the SM effective field theory extended 
with $a$~\cite{Gripaios:2016xuo}.) One can easily estimate 
$\Gamma(B^+\to K^+ a a)\sim (\text{few}\,\text{GeV})^9 /\Lambda^8$, 
which is of the order of $10^{-11}$ provided $\Lambda \lesssim 1$ TeV. For this 
reason, we will also consider the channel $B^+\to D^+ a a$. It tests 
operators such as $\mathcal{O}\sim 1/\Lambda^4 
a^2 (\overline{u_R}\gamma_\mu b_R) (\overline{d_R}\gamma^\mu c_R)+\text{h.c.}$

We will assume that $a$ decays into muons with a 
width smaller than $\sim 10$~MeV and a lifetime shorter than $\sim 10$~fs. In 
this case, it will appear 
not to have any experimentally measurable flight distance and will appear to 
have  zero width. This is easily achieved if $a$ is muonphilic with 
$10^{-5}\lesssim g_{22}^e\lesssim 1$. The processes discussed so far leads 
therefore to four-muon 
final states with and without additional mesons and with the muons forming 
pairs of two identical masses.

\subsection{Explicit model}
\label{sec:models}
Light scalars are natural within CHMs, for they are approximate Nambu-Goldstone 
Bosons (NGBs) arising from the spontaneous symmetry breaking 
$\mathcal{G}/\mathcal{H}$ in the confinement of a new strong sector 
at a scale $f\sim$ TeV. The simplest coset delivering the four Higgs degrees of 
freedom as well as a new scalar singlet $a$ is 
$SO(6)/SO(5)$~\cite{Gripaios:2009pe}. Interestingly, it can be UV completed in 
four dimensions~\cite{Ferretti:2016upr}.

In this model, the SM fermions do not couple directly to the Higgs boson. They rather mix with 
other composite resonances that do interact with the Higgs boson. Thus, the Yukawa 
Lagrangian depends on the quantum numbers of the aforementioned resonances.
As a simple yet 
realistic example, we assume that the second generation leptons mix with 
two fundamental representations $\mathbf{6}$ of $SO(6)$. An 
equivalent 
description is 
the 
embedding of the elementary leptons into incomplete fundamental representations 
of $SO(6)$. The most general such embedding depends on a single positive 
parameter $\gamma$ to give 
\begin{equation}
 \text{L}_L = (-\text{i}\mu_L, \mu_L, \text{i}\nu_L, \nu_L, 0, 0)~, 
\quad \text{L}_R = (0, 0, 0, 0, \gamma \mu_R, \mu_R)~.
\end{equation}
Using the corresponding Goldstone matrix
\begin{equation}
 \text{U}=\left[\begin{array}{cccc}
              1_{3\times 3} &  &  & \\
              &1-h^2/(f^2+\Pi) & -ha/(f^2+\Pi) & h/f\\
              &-ha/(f^2+\Pi)  & 1 -a^2/(f^2+\Pi) & a/f\\
              &-h/f & -a/f & \Pi/f^2
             \end{array}\right]~, \Pi = 
f^2\left(1-\frac{h^2}{f^2}-\frac{a^2}{f^2}\right)^{1/2}~,
\end{equation}
one obtains the leading-order Yukawa Lagrangian
\begin{align}
 L = &-\frac{y_\mu}{\sqrt{2}} f\overline{(U^T \text{L}_L)_1} (U^T 
\text{L}_R)_1 + 
\text{h.c.} = -\frac{1}{\sqrt{2}} y_\mu \overline{\mu_L} h\mu_R\left[ 1 + 
\frac{\gamma}{f}a + \cdots\right] + \text{h.c.}
\end{align}
The coupling $y_\mu\sim 6\times 10^{-4}$ stands for the muon Yukawa.
The subindex in $(U^T \text{L}_R)_1 $ indicates the projection of 
the fundamental representation of $SO(6)$ into the singlet of $SO(5)$ according 
to the decomposition $\mathbf{6} = 1+ \mathbf{5}$.
The ellipsis 
stands for terms containing higher powers of $a$.
Likewise, the one-loop induced potential for $a$ reads:
\begin{align}
 V &= c_R f^4 y_\mu^2 (U^T\Lambda_R)_1 
(U^T \Lambda_R)_1 + c_L f^4 y_\mu^2 (U^T\Lambda_L^I)_1 \nonumber
(U^T \Lambda_{I L})_1\\
&\sim c_R f^2 
y_\mu^2\left(\gamma^2-1\right)a^2+\cdots
\end{align}
where we have neglected terms not involving $a$. $\Lambda_R$ stands for L$_R/\mu_R$; analogously for $\Lambda_L^I$ with $I$ the flavour index.  $c_R$ is a free 
parameter encoding the details of the strong sectors. Its size can be estimated 
using naive power counting~\cite{Giudice:2007fh}, $c_R\sim 
g_*^2/(32\pi^2)$, with $1\lesssim g_*\lesssim \sqrt{4\pi}$ the typical coupling between resonances.
All in all, we obtain
\begin{align}
  m_a &\sim \frac{g_\rho y_\mu}{4\pi}  \sqrt{(\gamma^2-1)}\, f~,  \\
  g_{22}^e &\sim \frac{m_\mu}{f}\gamma~.
\end{align}
The scalar $a$ decays 
$100\,\%$ into muons. The other fermions respect this 
phenomenology provided they do not break the shift symmetry $a\rightarrow a+\text{constant}$, nor 
$a\rightarrow -a$. These two 
conditions can hold simultaneously if the left (right) chiralities mix with 
\textit{e.g.} $\mathbf{6}$ ($1$), $\mathbf{6}$ ($\mathbf{15}$) or $\mathbf{6}$ 
($\mathbf{20'}$).

The scalar defined above can explain the 
longstanding anomaly on the magnetic moment of the muon~\cite{Liu:2018xkx}. In this concrete model, we can fit
the experimental measurement $(\Delta a_\mu)_\text{obs} = (2.74\pm 0.73)\times 10^{-9}$~\cite{Bennett:2006fi}
within two standard deviations for $g_* = 2$ and $f = 800$ GeV and $\gamma\gtrsim 10$.
Fitting the experimental observation within one standard deviation is in principle possible, but it 
requires even  larger values of $\gamma$, too small values of $g_*$ (which 
would contradict the strongly coupled nature of the composite sector) and $f \lesssim 800$ GeV 
(which is in tension with Higgs and electroweak precision 
data~\cite{Ghosh:2015wiz}).
Therefore, the value $m_a \sim 
1$ GeV is a very likely value in  this setup.

New composite vector bosons $V$ explaining the observed 
anomalies in $R_{K^{(*)}}$ appear also naturally in this 
framework~\cite{Niehoff:2015bfa,Niehoff:2015iaa,Carmona:2015ena,Megias:2016bde, 
GarciaGarcia:2016nvr,Megias:2017ove,Sannino:2017utc,Carmona:2017fsn,
Chala:2018igk,Falkowski:2018dsl}. 
These particles decay preferably into composite states~\cite{Chala:2018igk}, 
being too broad 
and too heavy to be directly detected unless very dedicated LHC analyses are 
performed for masses $m_{V}\lesssim 3$ 
TeV~\cite{Chala:2018igk}.~\footnote{Irrespectively of $g'$, the interaction 
between $V$ and the heavy resonances $L$ triggers the decay $B_s^0\rightarrow 
V^*\rightarrow \mu^+ L^*, L^*\to a\mu^-$, where $^*$ denotes off-shellness. For 
couplings equal to the unit, the corresponding width at tree level reads 
exactly $\Gamma = f_B^2/(m_L^4 m_V^4) (1+m_a/m_B)^5 (1-m_a/m_B)^5 m_B^7$, with 
$m_L$ the mass of $L$. Even for $m_L \sim 500$ GeV and $m_V \sim 1$ TeV, the 
corresponding branching ratio is $\sim 10^{-13}$ and therefore beyond the reach 
of our analysis.}

The interaction with the SM fermions takes the form of Eq.~\ref{eq:Lagrangian}, 
with $g^d_{23}\sim 
0.002\, m_V^2/\text{TeV}^2$~\cite{DiLuzio:2017fdq,Chala:2018igk}.
Relying on these results, we consider for reference the benchmark point
\begin{equation}
 \text{BP}: m_a = 1~\text{GeV}~, ~ m_V = 4~\text{TeV}~, ~ g_{23}^d \sim 0.03~.
\end{equation}
For $g' \lesssim 0.1$, this point satisfies all current constraints from LHC 
searches and measurements of $\Delta M_s$. Constraints set by the latter 
observable could be competitive if the more recent predictions of the SM value 
are confirmed~\cite{Bazavov:2016nty,DiLuzio:2017fdq}.
However, we will show that a signal should be 
observable with the future upgrades of the LHCb experiment.
In particular, we obtain $\mathcal{B}(B_s^0\rightarrow \mu^+\mu^-\mu^+\mu^-)\sim 4\times 10^{-9}$, 
$\mathcal{B}(B^+\rightarrow K^+\mu^+\mu^-\mu^+\mu^-)\sim 10^{-9}$. Note that the 
final state with $K^+$ meson is of the same order of magnitude as the four body 
final state. Together with the fact that the $B^+$ meson has a higher production 
cross section 
than $B_s^{0}$ in $pp$ collisions, this suggests that $B^+\rightarrow 
K^+\mu^+\mu^-\mu^+\mu^-$ is a key signature to explore for this kind of models. This 
channel has not been experimentally explored though.
Similar conclusions were pointed out in Ref.~\cite{Nelson:2013ula} in the 
context of dark sectors\footnote{In this case, rare $B$ meson decays are
triggered by flavour-violating scalars.}.

\section{Reach of the LHCb}
\label{sec:BF}

LHCb has searched for the decays $B^0_{(s)} \to \mu^+ \mu^- \mu^+ \mu^-$~\cite{Aaij:2016kfs} and has set the limits \mbox{$\mathcal{B}(B^0 \to \mu^+ \mu^- \mu+ \mu^-) < 7 \times 10^{-10}$} and $\mathcal{B}(B^0_s \to \mu^+ \mu^- \mu+ \mu^-) < 2.5 \times 10^{-9}$ with a 3~fb$^{-1}$ dataset at the collision energies of $\sqrt{s} = 7$ and $\sqrt{s}=8$~TeV. However, there is a limitation in this analysis as it places a veto on the mass of the muon pairs to be close to the $\varphi$ or $J/\psi$ mass, to avoid background from the decay $B^0_s \to J/\psi \varphi$ followed by both vector mesons decaying to a pair of muons. As the $a$ mass is likely very close to the $\varphi$ mass, the current analysis is not sufficiently general. If a new analysis removes the veto around the $\varphi$ mass, and instead imposes a requirement that two opposite muon combinations recombine to the same invariant mass, the limit should stay the same and the background from $B^0_s \to J/\psi \varphi$ would still be avoided.

Due to the four muons in the final state for the LHCb analysis, the combinatorial background to a possible signal is extremely low. As an essentially background free analysis, even in the far future, the branching fraction limit can be expected to scale inversely with the number of $B$ mesons produced.

\begin{figure}[t]
 \begin{center}
 \includegraphics[width=0.49\columnwidth]{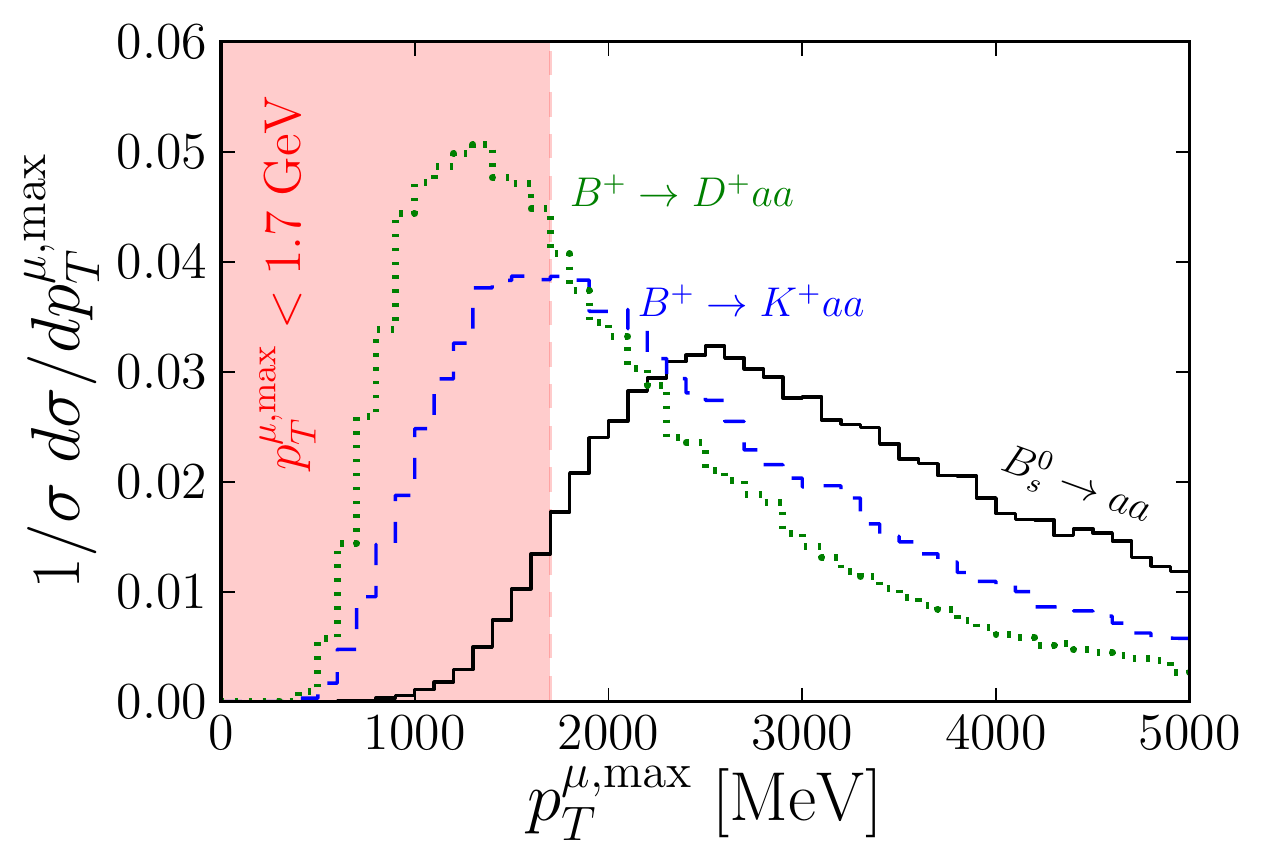}
  \includegraphics[width=0.49\columnwidth]{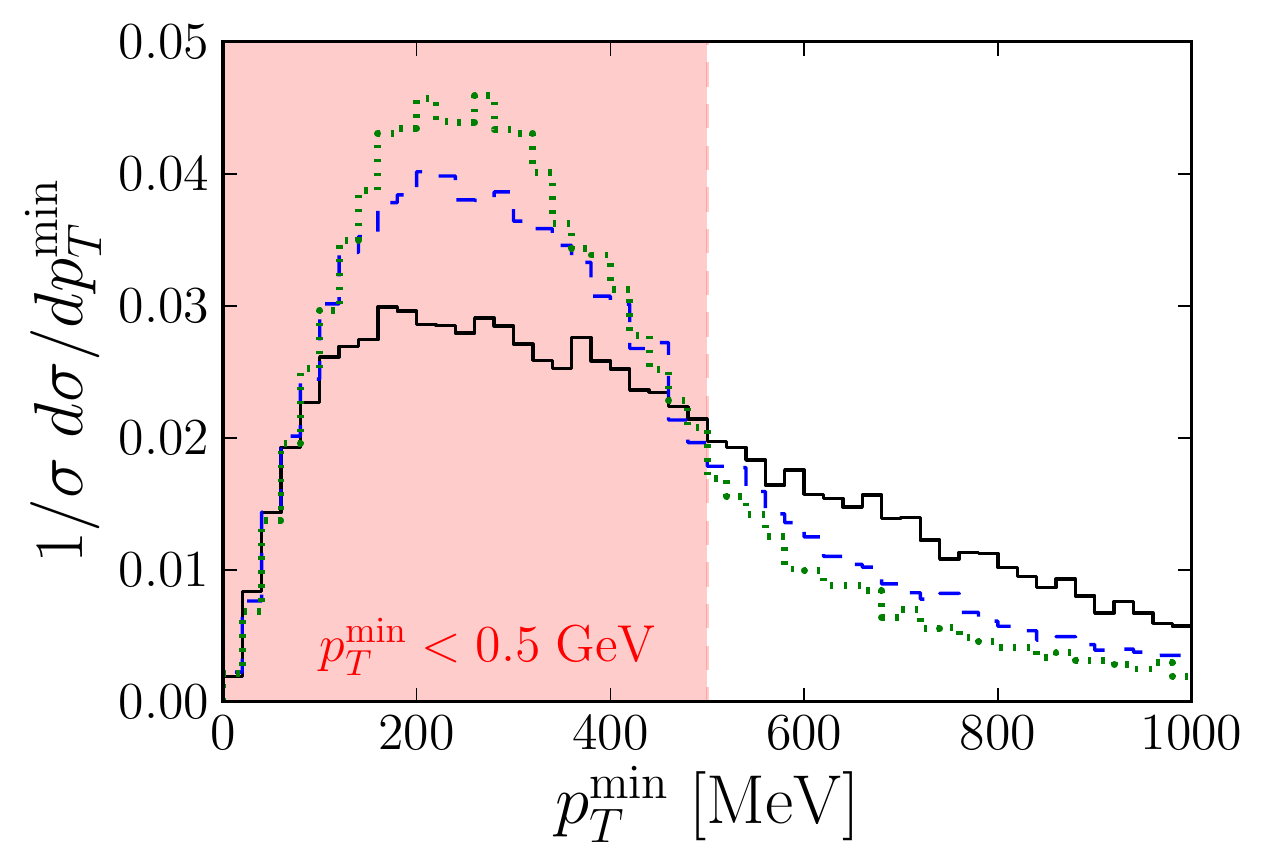}
  \includegraphics[width=0.47\columnwidth]{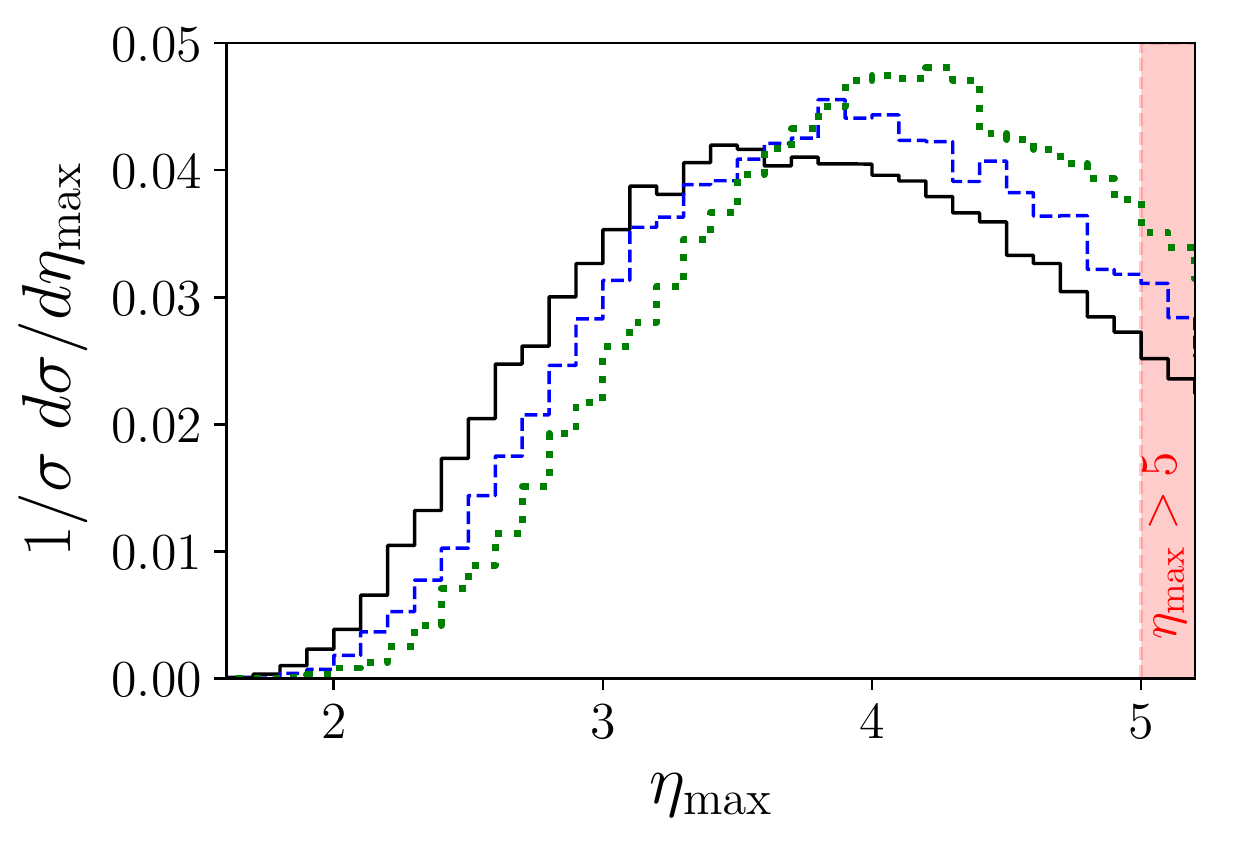}\hspace{0.35cm}
  \includegraphics[width=0.47\columnwidth]{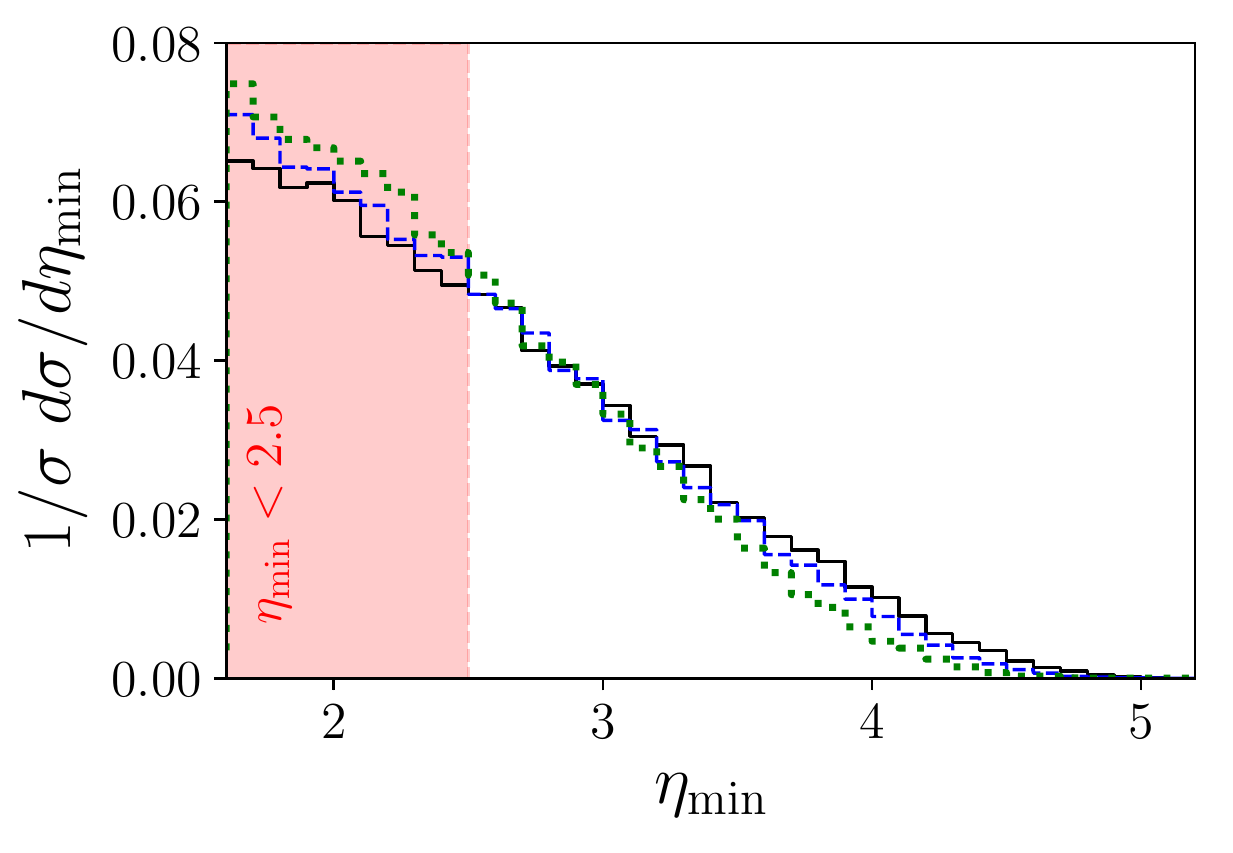}\hspace{0.1cm}
 \caption{\it Upper left) Normalised distribution of the transverse momentum of 
the hardest muon in the case $B_s^0\rightarrow aa$ (solid black), $B^+\to K^+ 
aa$ (dashed blue) and $B^+\to D^+ aa$ (dotted green). Upper right) Same as 
before but for the transverse momentum of the softest particle. Bottom left) 
Same as before but for the pseudorapidity of the most central particle. Bottom 
right) Same as before but for the pseudorapidity of the most forward particle.}
\label{fig:distributions}
 \end{center}
\end{figure}

In the run periods of Upgrade-I and Upgrade-II of LHCb, the collision energy 
will be $\sqrt{s}=14$~TeV. As the $b$ cross-section is scaling more or less in 
direct proportion to the collision energy, the amount of $B$ mesons produced per 
inverse fb, can be expected to be around a factor $14/7.7=1.8$ higher compared 
to the average Run-1 conditions of the LHC. Expectations below will be quoted 
for the end of LHCb (9~fb$^{-1}$), end of Upgrade-I (50~fb$^{-1}$) and end of 
upgrade-II (300~fb$^{-1}$). The naive scaling factors, compared to the current 
Run-1 for these, and taking the different $b$ cross-sections into account, are 
for LHCb a factor 4.4, for Upgrade-I a factor 29, and for Upgrade-II a factor 
180. Thus for the limits on $\mathcal{B}(B^0_s \to \mu^+ \mu^- \mu+ \mu^-)$ we 
should expect $6 \times 10^{-10}$, $9 \times 10^{-11}$ and $1.4 \times 
10^{-11}$, respectively. This assumes no changes to the trigger or tracking 
performance in the upgrades of LHCb.

Irreducible backgrounds to the decay have to be considered. The decay $B^0_s \to \varphi\varphi$ with $\varphi \to \mu^+\mu^-$ is one of these. Using the measured branching fractions~\cite{Aaij:2015cxj,Ambrosino:2004vg} we get
$\mathcal{B}(B^0_s \to (\varphi \to \mu^+\mu^-) (\varphi \to \mu^+\mu^-)) = 1.84 \times 10^{-5} \times (2.89 \times 10^{-4})^2 = 1.5 \times 10^{-12}$. As can be seen from the expected limits above, even at the end of LHCb Upgrade~II, this is not relevant. For the equivalent decay mode of the $B^0$, the measured branching fraction limit for the $B^0 \to \varphi \varphi$ decay is three orders of magnitude below the $B^0_s$ mode and thus even less of a concern. The decay $B^0_s \to \varphi \mu^+ \mu^-$ has a measured differential rate of $2.6 \times 10^{-8}$~GeV$^{-2}$ in the region of the squared dimuon mass close to the $\varphi$ mass~\cite{Aaij:2015esa}. Letting the $\varphi$ decay to a muon pair and considering a mass region with width of around 20 MeV, corresponding to a realistic mass resolution, this will give a background at the $10^{-13}$ level and is thus not relevant.

A simplified model for which limits the LHCb experiment can be made for similar decay modes. When 
comparing different $B$ hadrons, the relative production fractions as measured 
at $\sqrt{s}=7$~TeV collisions in the LHCb acceptance~\cite{Aaij:2011jp} are taken into 
account. The relative production fractions are not expected to change 
significantly with collision energy as they are determined from the 
fragmentation process. From this, 
we conclude that the production of 
$B^+$ 
and $B^0$ are the same 
and that the production of $B^0_s$ mesons is a factor 3.7 less common. For the 
reconstruction in LHCb, it is assumed that the efficiency is 95\,\% per track 
inside the fiducial volume defined by the pseudorapidity $2.5 < \eta < 5.0$ and 
that tracks have a transverse momentum above $0.5$~GeV with respect to the beam 
axis to be reconstructed. For the trigger it is assumed that at least one reconstructed muon should 
have a transverse momentum above $1.7$~GeV. The effect of these criteria is that 
final states with a larger number of particles have a lower efficiency, both due 
to the requirement that all tracks have to be reconstructed but also due to that 
the muons turn softer and the trigger efficiency thus is getting lower; see Fig.~\ref{fig:distributions}.

For the positive identification of muons, it is assumed that the
efficiency is 100\,\% for muons with a total momentum above
2.5~GeV. It is assumed that no or only very loose particle
identification is required on the charged hadrons. For the $D^{+}$
reconstruction, it is assumed that only the
$D^{+} \to K^{-}\pi^{+}\pi^{+}$ final state is used. This is the
easiest decay mode to reconstruct and has a branching fraction of
$9.0\,\%$~\cite{PDG2018}. The final state with a semileptonic decay of
the $D^+$ could also be considered, in an analysis similar to the
$B^+ \to \mu^+\mu^-\mu^+\nu$~\cite{Aaij:2018pka} analysis carried out
by LHCb. However, to estimate the reconstruction efficiency of this
five charged lepton final state with a neutrino would require a full
detector level simulation which is beyond this paper. For the $K^{*0}$
case, we consider the decay into $K^+ \pi^-$, which has a branching
ratio of $\sim 67\,\%$.

All overall efficiencies 
for a given final state are evaluated relative to the published analysis on the 
decays $B^0_{(s)} \to \mu^+ \mu^- \mu^+ \mu^-$~\cite{Aaij:2016kfs}. As the 
trigger and main selection are the same for all these decays, this provides a 
robust normalisation method.

Simulations are carried out using \texttt{Pythia~8}~\cite{Sjostrand:2007gs} for the production of $B$ mesons in $pp$ collisions and \texttt{EvtGen}~\cite{Lange:2001uf} for the subsequent decays. The decays are assumed to be of the type $B \to (M)aa$ with $a \to \mu^+ \mu^-$ and with $M$ a possible meson in the final state. The $B$ meson decay is simulated with a flat phase space distribution. If the hadron is unstable, it is decayed to stable particles using the default model in EvtGen and with branching fractions taken from the PDG~\cite{PDG2018}. A summary of the expected limits that can be set are given in Table~\ref{tab:limits}.

\begin{table}
  \centering
  \resizebox{0.8\textwidth}{!}{\begin{tabular}[c]{|l|c|c|c|}\hline
    \textbf{Decay} & \textbf{LHCb} & \textbf{Upgrade} I & \textbf{Upgrade II} \\
    \hline
    $B^0_s \to \mu^+ \mu^- \mu^+ \mu^-$   & 60 & 9 & 1.4 \\
    $B^0 \to \mu^+ \mu^- \mu^+ \mu^-$     & 15 & 2.3 & 0.4 \\
    $B^+ \to K^+ \mu^+ \mu^- \mu^+ \mu^-$ & 37 & 5 & 0.9 \\
    $B^0 \to K^{*0} \mu^+ \mu^- \mu^+ \mu^-$ & 100 & 16 & 2.7 \\
    $B^+ \to D^+ \mu^+ \mu^- \mu^+ \mu^-$ & 1300 & 200 & 32\\\hline
  \end{tabular}}
  \caption{Expected branching fraction limits for the different decays
    under consideration in units of $10^{-11}$. It is assumed that the
    $D^+$ meson is only reconstructed in the $K^-\pi^+\pi^+$ final
    state.}
  \label{tab:limits}
\end{table}

\begin{figure}[t]
 \begin{center}
 \includegraphics[width=0.49\columnwidth]{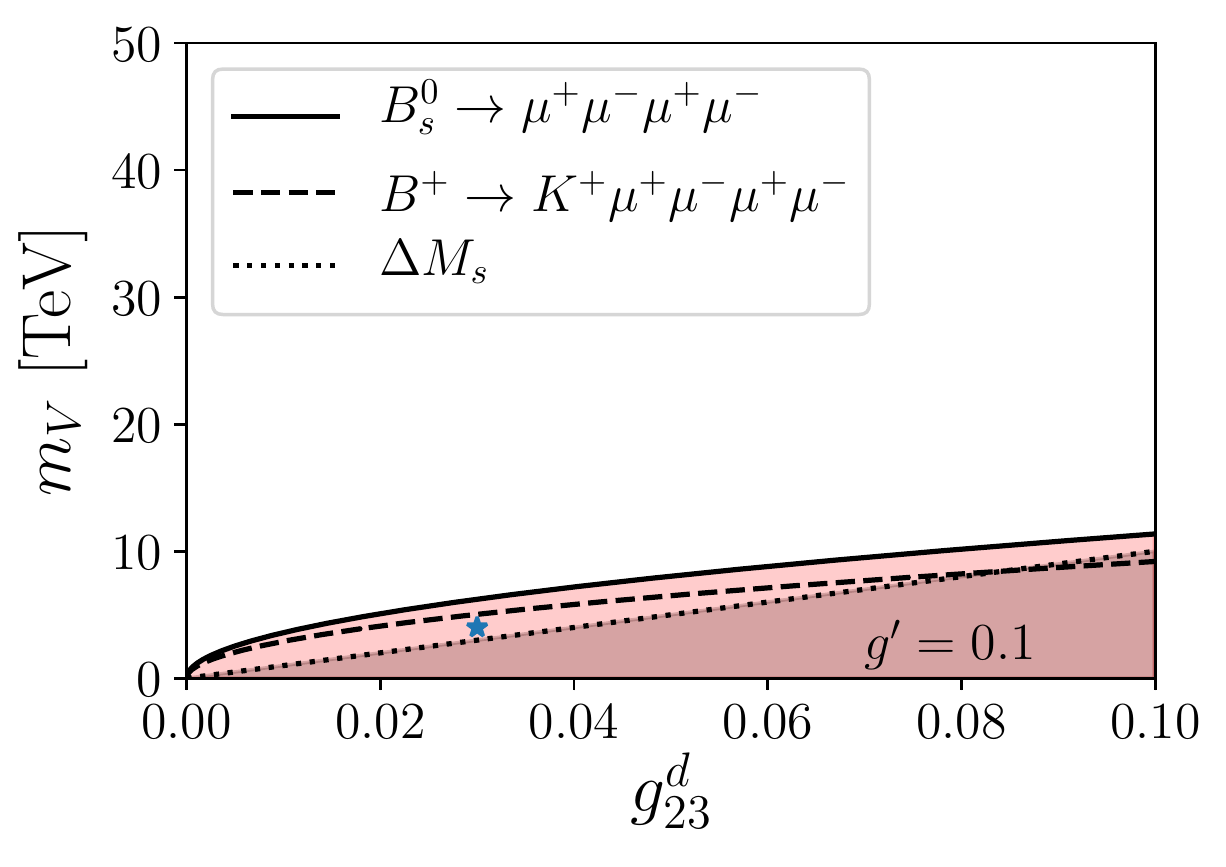}
  \includegraphics[width=0.49\columnwidth]{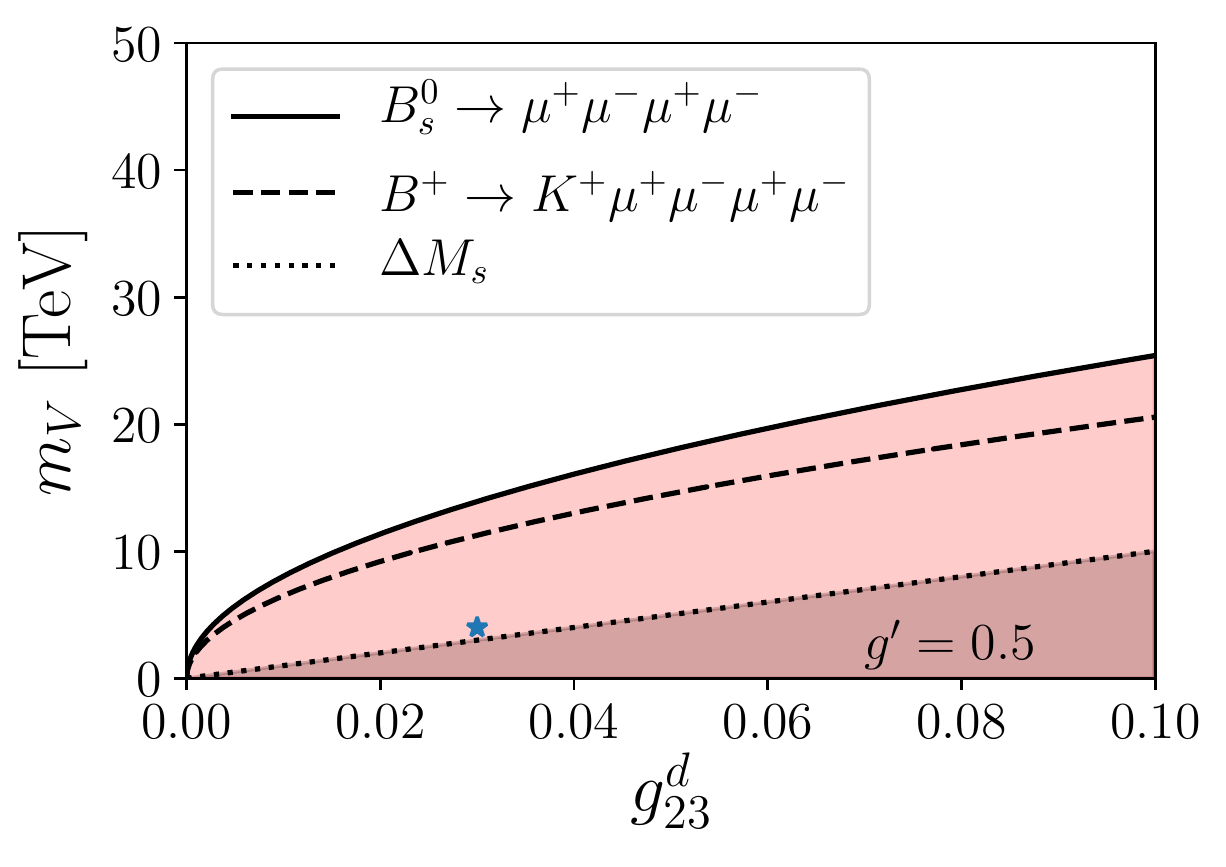}
  \includegraphics[width=0.49\columnwidth]{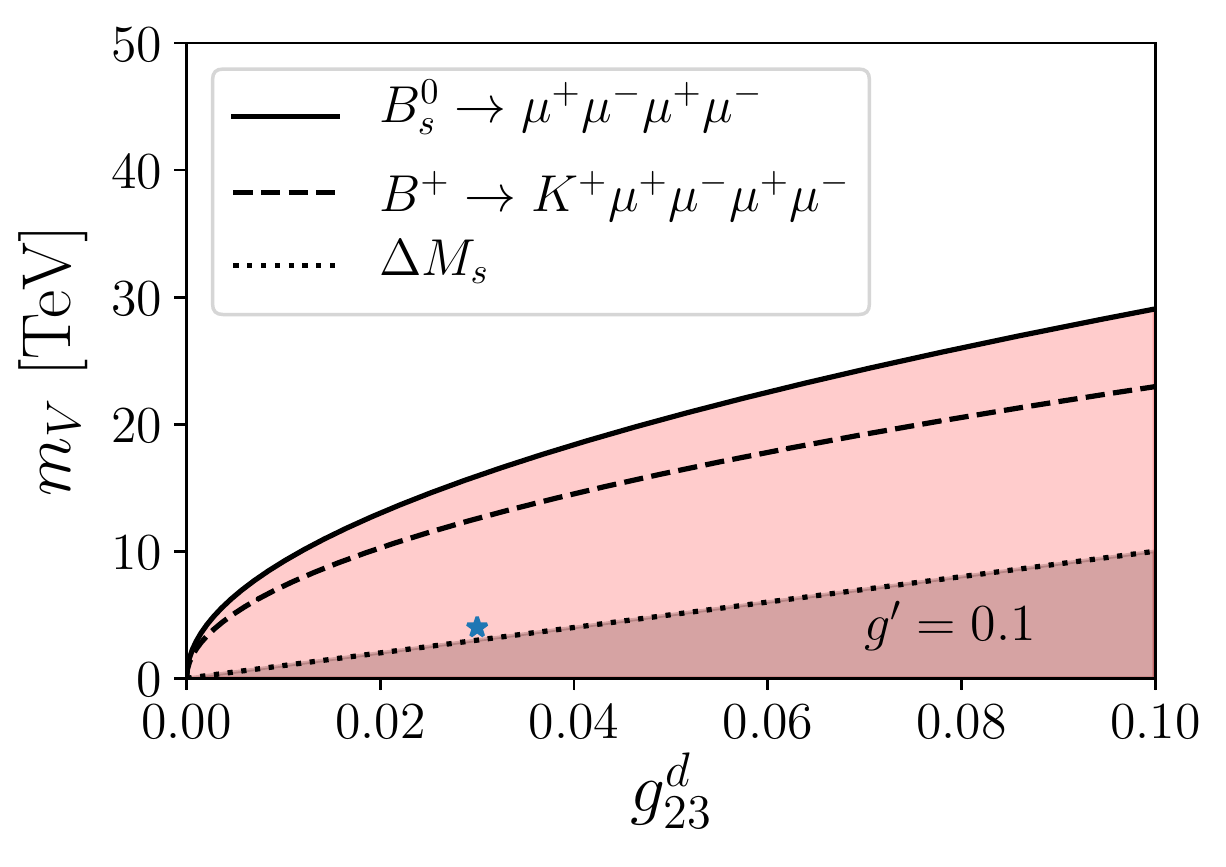}
  \includegraphics[width=0.49\columnwidth]{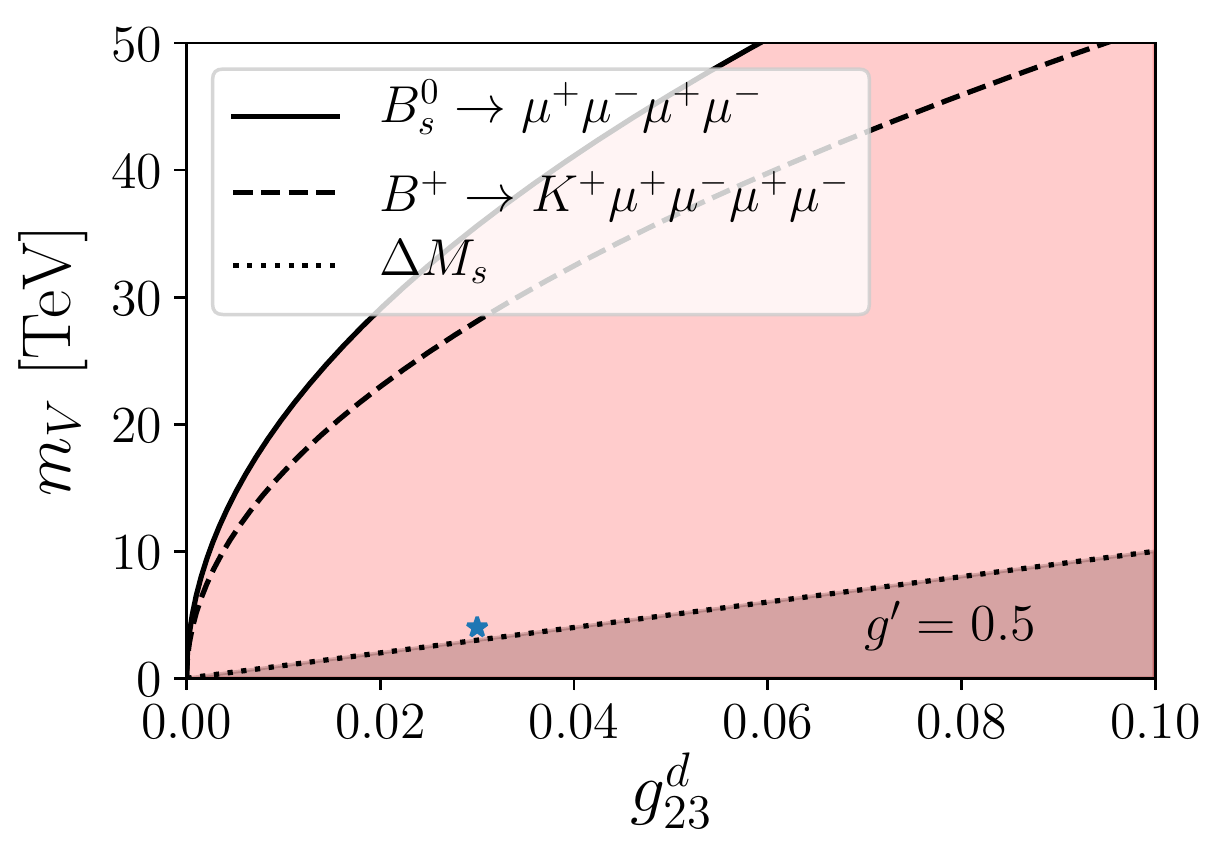}
 \caption{\it Upper left) Region in the plane $(g_{23}^d, m_V)$ that can be 
tested at the current run of the LHCb (light) for $g' = 0.1$ versus the area 
excluded by measurements of $\Delta M_s$ (dark). Upper right) Same as before but 
for $g' = 0.5$. Bottom left) Same as before but for the LHCb Upgrade II and 
$g'=0.1$ Bottom right) Same as before but for $g'=0.5$. 
In all cases, the benchmark point defined in section~\ref{sec:models} is shown 
with a star for reference.}
\label{fig:final}
 \end{center}
\end{figure}

Translated to the plane $(g_{23}^d, m_V)$ for given values of $g'$, the limits from $B_s^0\to \mu^+\mu^-\mu^+\mu^-$ and $B^+\to K^+\mu^+\mu^-\mu^+\mu^-$ are compared with those from $\Delta M_s$ in Fig.~\ref{fig:final}. Interestingly, we see that scales of several tens of TeV not yet probed by current experiments could be tested in the Upgrade II of the LHCb with our analysis.

\section{Conclusions}\label{sec:conclusions}

We have considered scenarios involving new heavy and flavour-violating vectors 
$V$ as well as light scalars $a$. 
We have shown that these particles give rise to rare $B$ 
meson decays that are not yet probed. As the preferred mass of the scalar $a$ 
lies 
inside the window vetoed by current LHCb searches, namely $[950,1090]$ MeV, even 
the simplest decay mode  $B^0_{(s)}\to \mu^+\mu^-\mu^+\mu^-$ is not fully 
probed. Other decay modes of interest are 
$B^+\to K^+(D^+) 
\mu^+\mu^-\mu^+\mu^-$ and $B^0\to K^{*0}\mu^+\mu^-\mu^+\mu^-$. We have shown 
that the five-body final state can be as significant as the four-body. Relying 
on simulations, we have estimated the reach of 
the LHCb experiment for these processes in the current run and in Upgrades I and 
II. In the Upgrade II scenario we expect that branching fraction limits in the 
$10^{-11}$ region can be reached.

Finally, we emphasize again that the decays into a meson and $aa$ are the 
only 
sensible probe of different effective operators in the SM effective-field 
theory extended with a scalar singlet.
We therefore encourage the experimental collaborations to consider these processes in future analyses.

\vskip 1 \baselineskip

\noindent {\textit{Acknowledgments}.  We would like to thank Martin Bauer and 
Jakub 
Scholtz for helpful discussions. MC is 
supported by the Royal Society under the Newton International Fellowship 
programme. MS was 
supported by the Humboldt Society during the finalisation of parts of this 
work.}

  
\bibliographystyle{style}
\bibliography{references}

\end{document}